\documentstyle[12pt]{article}
\begin{document}

\begin{center}
{\Large\bf India-based Neutrino Observatory\footnote{Talk presented at 
NuFact 03, 5$^{th}$ International Workshop on Neutrino Factories \& 
Superbeams, 5-11, June 2003, Columbia University, New York.} }
\vskip .2cm
{\bf G.Rajasekaran} \\
{\it Institute of Mathematical Sciences} \\
{\it Madras-600 113,India}
\end{center}
{\bf Abstract:}{\sf An introduction to India-based Neutrino Observatory
and a brief status report are presented.The two possible sites are
described along with their special advantages.The proposed detector and
its physics capabilities for atmospheric neutrinos and long-base-line
experiments are discussed.}

\paragraph{Introduction:}
Historically,the Indian initiative in Cosmic Rays and Neutrino Physics
experiments goes back several decades.In fact,atmospheric neutrinos were
first detected in the Kolar Gold Fields (KGF) experiments in India almost
4 decades ago. KGF were one of the deepest mines in the world.When the
cosmic ray muon experiments were set up at deeper and deeper levels in the
mines, the counters fell silent at a particular depth.It was realised that
at those depths and beyond, atmospheric neutrinos could be detected. They
went ahead and detected them.That was in 1965 and it was the beginning of
atmospheric neutrino physics.However, the deeper levels of KGF are now
closed.

Sometime ago it was decided to revive neutrino experiments in India and a
major collaboration involving about 12 institutions has been formed.This
is the India-based Neutrino Observatory (INO) project.More than 60
scientists have already joined and feasibility studies are in progress.
Two possible sites for the underground laboratory have been located. A
magnetised tracking iron calorimeter of 30-50 Kton with RPC detector
elements is under design and prototyping.In the first stage the aim will
be to study atmospheric neutrinos and in the next stage this detector is
envisaged as the far detector for a long-base-line neutrino
experiment.International collaboration is invited.

\paragraph{A Tale of Two Sites:} 
We give more details about the two sites:
PUSHEP (Lat 11.5 deg N, Long 76.6 deg E): Under the Nilgiri Mountains
in South India; adjacent to a hydel project PUSHEP (Pykara Ultimate
Stage HydroElectric Project); vertical overburden in the range 1.3-
1.4 Km and all-around cover of more than 1 Km; laboratory cavern to be 
dug at the end of a tunnel of length about 2 Km;located in the Southern
Peninsular Shield; uniform rock medium of mean density 2.8 gms/cc;
seismic zone 2;close to the Cosmic Ray Laboratory of TIFR and the
Radio Astronomy Centre of TIFR,both in the hill station Ooty; close
to big cities (with airports) like Coimbatore and Bangalore,with
excellent industrial and academic infrastructure.Detailed survey
of the region is complete.

RAMMAM (Lat 27 deg N, Long 88 deg E):Under the Himalayas, in the
Darjeeling District of West Bengal; a tunnel of 3-5 Km can reach an
overburden of 1.5-1.85 Km or even more; seismic zone 4; detailed
survey is in progress.

Both sites are excellent for an underground neutrino laboratory. After a
critical evaluation of the relative advantages and disadvantages of both
sites with respect to the physics goals and practical aspects, one of them
will be chosen.

\paragraph{Detector:}
To start with,it is proposed to build a magnetised tracking iron
calorimeter,based on Monolith design.It will consist of 140 layers of 6 cm
thick iron plates interleaved with 2.5 cm air gap containing the active
detector elements.The dimensions will be 15 m X 32 m X 12 m (ht) and the
weight about 30Ktons.The detector elements will be glass RPC's (Resistive
Plate Chambers),with nanosecond timing, to provide up-down
discrimination.The detector will be sensitive to muons and other charged
particles.

A magnetic field of 1-1.3 Tesla will be an important feature of
the detector.This will provide efficient energy-momentum
resolution and also charge-identification which is an essential
requirement for a far-end detector in a long-base-line experiment.

The detector will be constructed in a modular fashion,so that
additional modules can be added in future, to augment its capability
for the long-base-line experiment and other experiments. Emulsion sandwich is
also being considered as a possibility with the aim of detecting
the tau leptons.

\paragraph{Physics possibilities at INO:}
From the long term point of view,a neutrino detector located in India will
have several advantages.A solar neutrino detector at a low latitude
(PUSHEP at 11.5 deg) will detect solar neutrinos passing through the core
of the Earth.A geoneutrino detector at RAMMAM will detect the geoneutrinos
from the unusually thick continental crust below the Tibetan plateau.Very
long base lines from neutrino factories become possible with detectors
located in India.The baseline lengths from a neutrino factory at
JHF(Japan) to PUSHEP or RAMMAM are 6,595 Km (31 deg) and 4,880 Km (22 deg)
respectively,the brackets showing the required dip angle of the muon decay
pipe. The corresponding numbers from a neutrino factory at CERN are 7,145
(34) and 6,890 (33) and from a factory at Fermilab are 11,300(62) and
10,500(55). The importance of multiple number of long base lines for
neutrino physics is now well recognised.Some of the above distances are
close to the "magic" baseline length of 7,200 Km.The very long base line
of 11,300 Km passes through 3000 Km of Earth's core and hence is likely to
play a major role in future neutrino tomography of the Earth.

In phase I, INO will be studying atmospheric neutrinos.The aim will be to
establish neutrino oscillations by showing the rise in the neutrino flux
after the minimum and to reduce the present uncertainty in the value of
$\Delta m^2_{32}$.In phase II, INO can play the role of the far detector
in a long-base-line experiment.The major tasks in this phase would be
probing $\theta_{13}$,determination of the sign of $\Delta m^2_{32}$ and
gaining a first glimpse of the CP violating phase $\delta$. 

We shall now show a few results of preliminary calculations by Raj Gandhi
and Anindya Datta on the physics capabilities of INO both in phase I and
phase II.The inputs used are muon detection threshold of 2 GeV and muon
energy resolution of 5 percent.All measurements in phase II involve wrong
sign muon detection and so backgrounds are low.

Fig 1 shows the up/down ratio of atmospheric nu as a function of L/E for
200 Kton-yr of operation of the INO detector. It is clear that
oscillations can be established.

Fig 2 gives the reach of $\sin\theta_{13}$ for nu factory
experiments as a function of the muon detection threshold energy. The
reach is defined as the value that will yield 10 signal events (wrong sign
muons) for a given kT-yr exposure. We have shown the plots for
Japan-RAMMAM and Fermilab-PUSHEP baselines. 

Fig 3 shows the
number of wrong-sign muon events as a function of $\Delta m^2_{23}$ for
three base lines from a neutrino factory in Japan, the detector being at
Beijing,Rammam or PUSHEP.The sign discriminating capability for either of
the two sites Rammam or PUSHEP is clearly demonstrated.

Because of lack of space we shall not show the graphs for the CP violating
phase, but calculations have been done for the ratio of wrong-sign muon
events for a run with negatively charged muons in the storage ring to that
for a run with positively charged muons.Although the CP violating effect
is weak for the Japan-PUSHEP baseline, it is clearly present for the
Japan-Rammam baseline. 

Finally,it must be emphasised that we need the support and cooperation of
neutrino enthusiasts all over the world.International collaboration for
the INO project is invited.Naba Mondal (nkm@tifr.res.in) is the
spokesperson for the project. More information on INO is available at the
INO web-site $\langle \langle$  www.imsc.res.in/$\sim$ino $\rangle \rangle$.

\begin{figure}[htp]
\vskip 11.0truecm
{\includegraphics{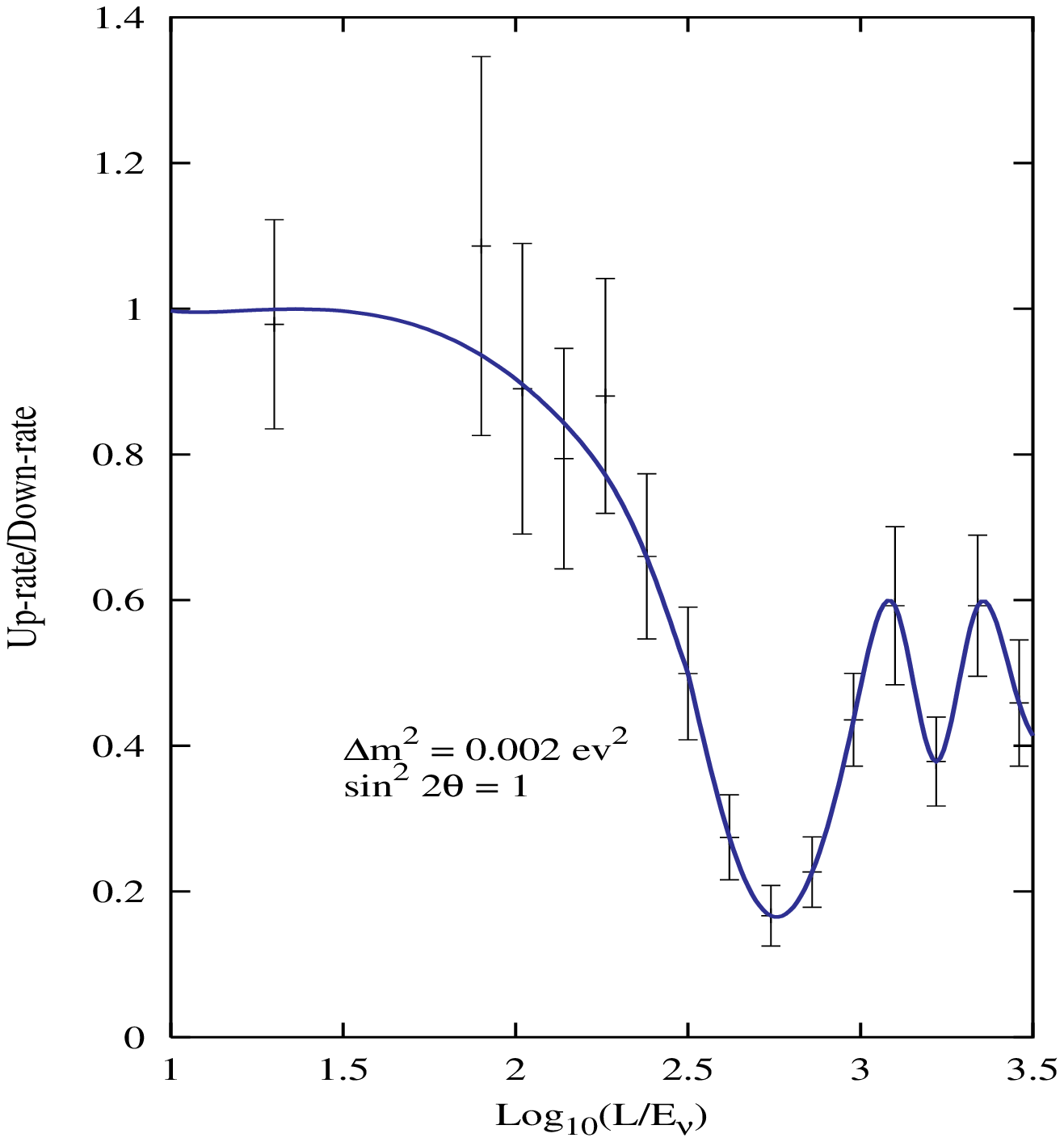}}
\caption{The up/down ratio of atmospheric $\nu$ vs L/E for
200 Kton-year.}
\end{figure}

\begin{figure}[htp]
\vskip 9.0truecm
{\includegraphics{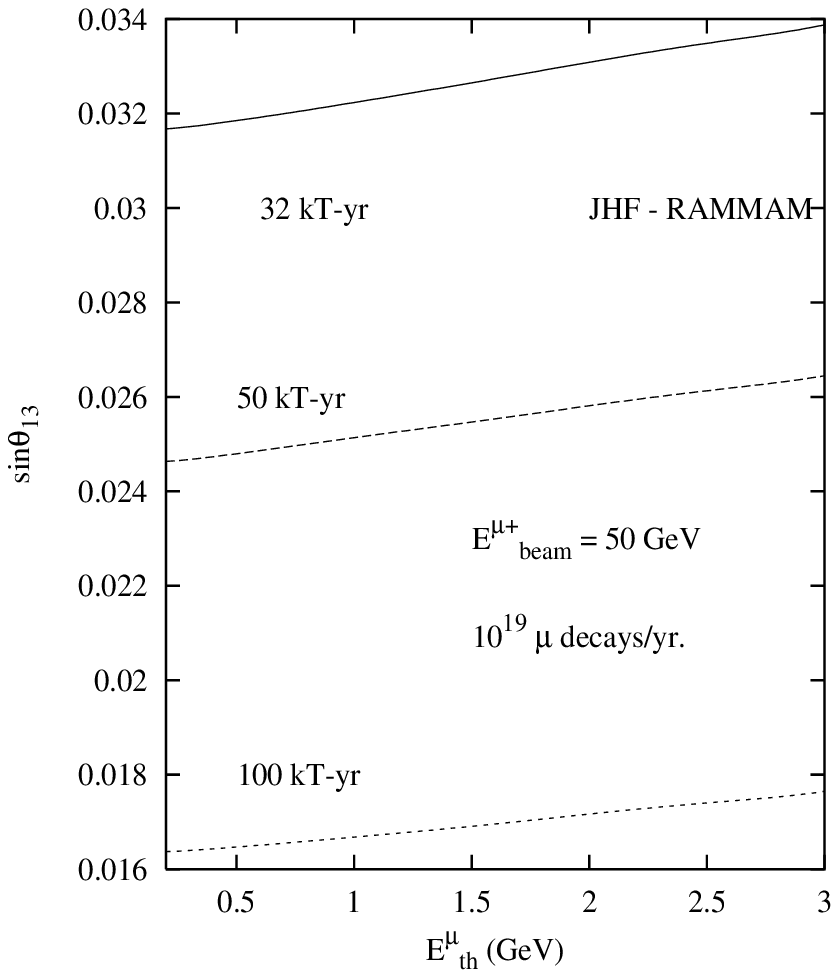}}
{\includegraphics{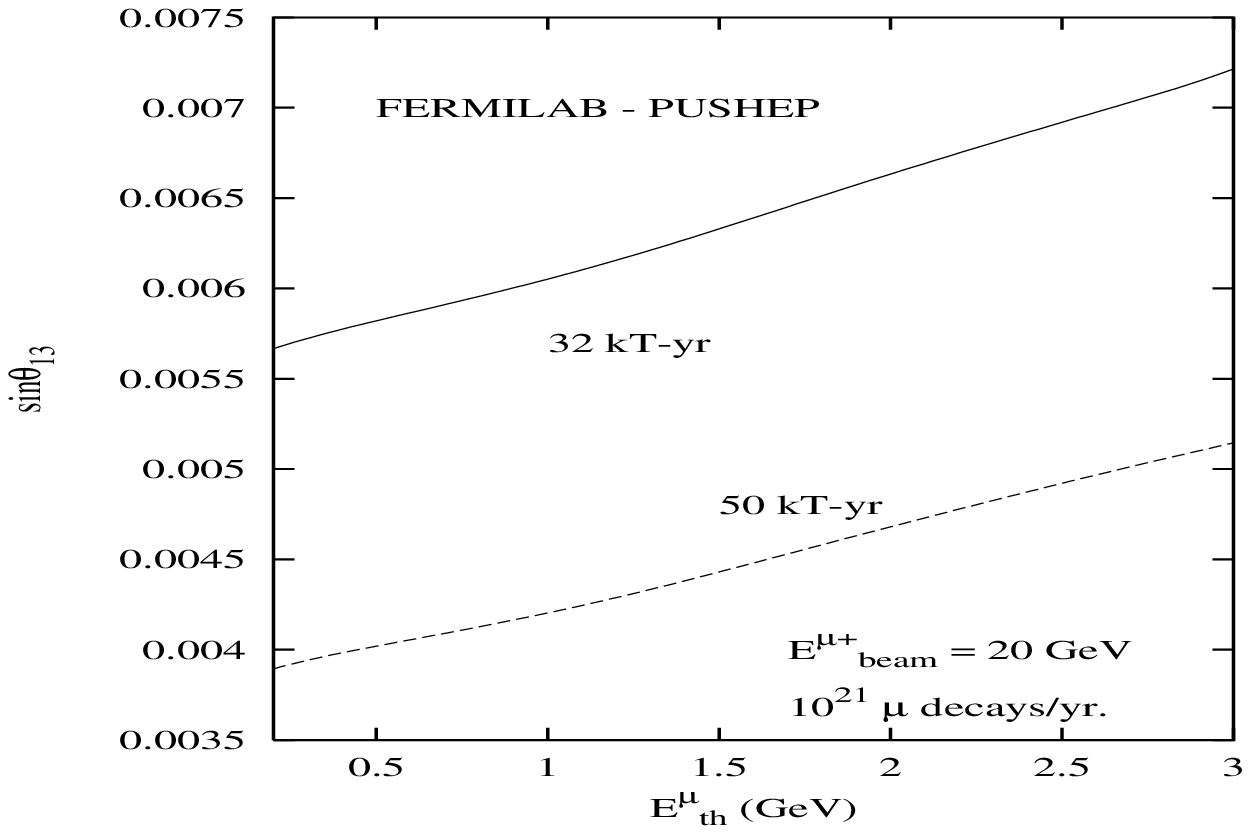}}
\caption{
$\sin\theta_{13}$ reach as a function of the muon threshold energy.
Left panel is for JHF to Rammam baseline. Right panel is for Fermilab to 
PUSHEP baseline.
}
\end{figure}

\begin{figure}[htp]
\vskip 10.0truecm
{\includegraphics{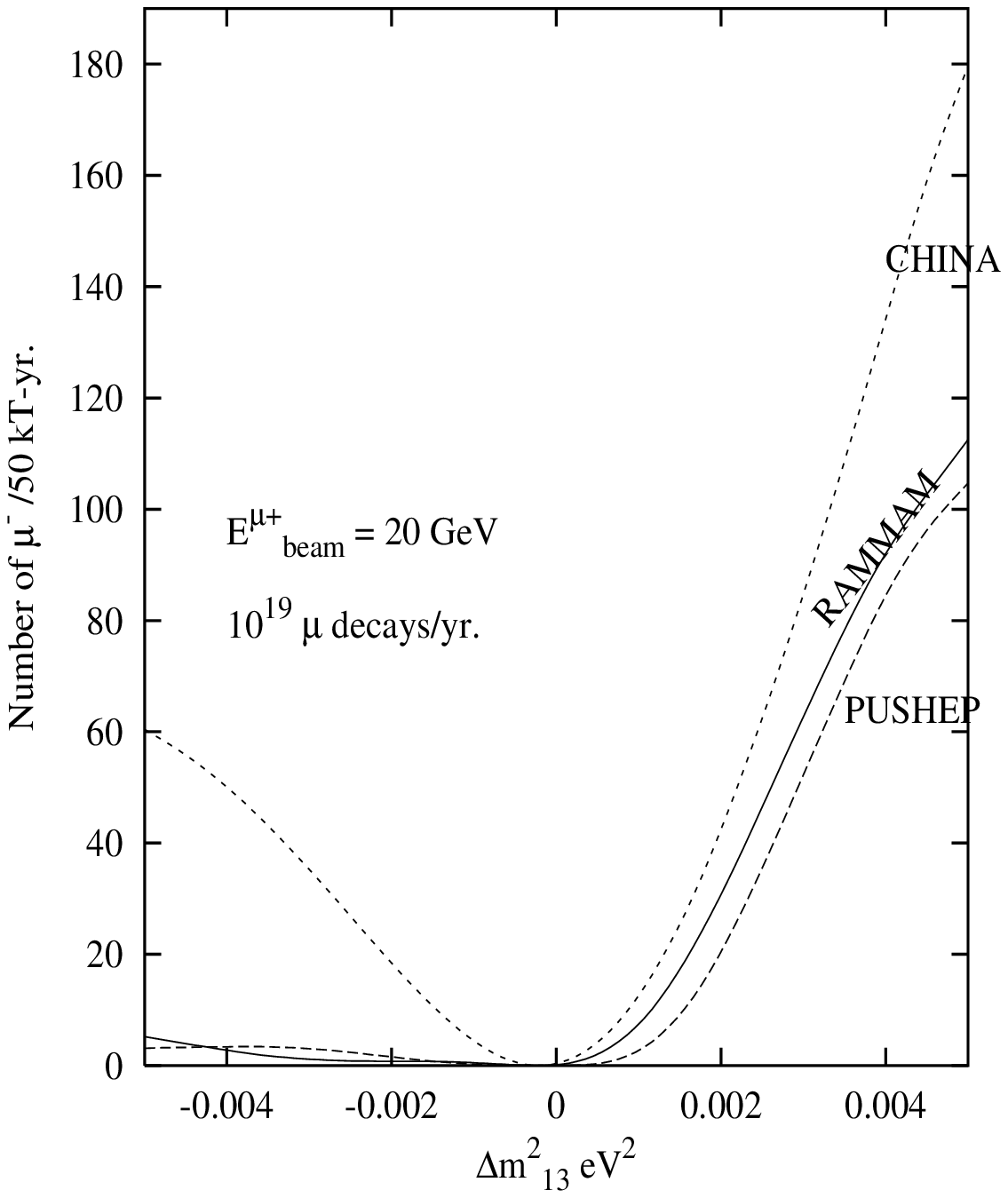}}
\caption{
The number of wrong-sign muon events vs $\Delta 
m^2_{23}$ corresponding to baselines from JHF to Beijing, Rammam and 
PUSHEP.}
\end{figure}

\end{document}